# Mn-doping induced magnetic properties in $Co_2CrO_4$ system

H. G. Zhang, W. H. Wang*, E. K. Liu, X. D. Tang, G. J. Li, J. L. Chen, H. W. Zhang, G. H. Wu

State Key Laboratory of Magnetism, Institute of Physics and Beijing National Laboratory for Condensed Mater Physics, Chinese Academy of Sciences, Beijing 100190, China

* Corresponding author, email: wenhong.wang@aphy.iphy.ac.cn

**Abstract**

Abundant phenomena in $CoCr_{2-x}Mn_xO_4$ (x = 0 ~ 2) samples such as magnetic compensation, magnetostriction and exchange bias effect have been observed and investigated in this work. A structure transition from cubic to tetragonal symmetry has been found in the samples with x around 1.4. It has shown that the doped $Mn^{3+}$ ions initially occupy the A (Co) sites when x < 0.2, and then mainly take the $B_1$ (Cr) sites. This behavior results in a role conversion of magnetic contributors, and thus a magnetic compensation between two competitively magnetic sublattices at the composition near x = 0.5. Furthermore, temperature compensation has also been found in the samples with x = 0.5 and 0.6, with the compensation temperature in the range of 45 ~ 75 K. The Mn-doping also changes the frustration degree and modulates the exchange interaction in this system, and thus prevents the formation of long range conical order of spins. Therefore, the magnetoelectric transition temperature at 23 K in $CoCr_2O_4$ is shifted to lower temperature with increased dopants. The magnetostriction effect in this Cobalt spinel system has been considered for the first time. The strain has a maximum value of about 240 ppm at x = 0.2 and shows the similar tendency as the compensation behaviors. Additionally, the exchange bias effect observed in the samples with x < 0.5 shows a negative value under low cooling field for x = 0.5.

# 1 Introduction

Multiferroic materials have recently attracted great attention due to their physical interest and potential applications[1-3]. These materials possess two or more ferroic orders, in which the one with coupling between the electric and magnetic degrees of freedom offers a wealth of opportunity for information storage applications[4]. So far, there are many kinds of single phase multiferroic materials have been found, which belong to several groups based on their microscopic origins[5,6]. One of them is called as frustrated magnets which exhibit magnetoelectric properties due to the ferroelectricity induced by the spiral spin structure[7-11]. As the first multiferroic material with both spontaneous magnetization and ferroelectric polarization of spin origin[12], $CoCr_2O_4$ which shows a unique conical-spiral ferrimagnetic spin order[13] has been intensively investigated recently.

The normal spinel type $CoCr_2O_4$ that with $Co^{2+}$ ions on the tetrahedral (A) sites and magnetic $Cr^{3+}$ ions on the octahedral (B) sites undergoes a ferrimagnetic transition at $T_C$ =93 K[13]. Thereafter, the system develops a spiral component short-range order which transforms into long-range order at 26K ($T_s$)[12,14]. The spiral plane on which the rotating component of spin lies is the (001) plane, while the spontaneous magnetization directs along the [001] or equivalent directions. The conical order of spins on $Cr^{3+}$ sublattice is thought to be the dominant origin of the ferroelectric polarization in $CoCr_2O_4$[12]. There are, however, some evidences that the dipole moments of $Co^{2+}$ ions on A-sites should not be neglected, especially at low temperature[15,16]. It is well known that the magnetic properties of spinel oxides are very sensitive to the nearest-neighbor arrangement of the cations and the isotropic antiferromagnetic exchange interactions between A-B and B-B sites. Therefore, if the cations on A/B sites are artificially modulated, there would be certainly some changes on the conical order (cone angle, for example) and the multiferroic properties relied on them[14,17]. Thus, doping on A/B sites is an effective way to modulate the magnetic and magnetoelectric properties of $CoCr_2O_4$.

Many previous works have concerned the substitution of cations in $CoCr_2O_4$ system[18-21], which mainly focused on the A-sites and only a few of them reported the results when $Cr^{3+}$ ions are substituted by another element[19,22]. Considering the Jahn–Teller effect of $Mn^{3+}$ ions on B-sites and its larger spin moment than $Cr^{3+}$, $Mn^{3+}$ substituted $CoCr_2O_4$ system may present large magnetization and interesting magnetic as well as multiferroic properties. In this work, a series of Mn doped $CoCr_2O_4$ samples have been synthesized and the systematic investigations about the structural, magnetic and magnetostrictive properties have been performed. It is found that the distribution of doped $Mn^{3+}$ ions results in apparent composition and temperature dependence compensation effects. The observations of magnetostriction and exchange bias effect in the series have also been reported in the present work.

# 2 Experimental details

A series of polycrystalline Mn-doped cobalt chromite samples with compositions of $CoCr_{2-x}Mn_xO_4$ (x=0-2) had been prepared in air by the solid-state reaction method using $Co_3O_4$ (99.7%), $Cr_2O_3$ (99%), and $MnO_2$ (97.5%) powders as precursors. The initial mixtures were well ground, pelletized, sintered twice at 1350/1250℃ for 24h, and were subsequently furnace cooled to room temperature.

Phase quality and structure of the samples were examined by powder XRD (X-ray diffraction) measurements using CuKa radiation (Rigaku Co., RINT-2400) with an angle (2θ) step of 0.02 between 2θ = 10° and 90°. Lattice parameters were obtained from cell refinement of the x-ray data using the Jade software. The extend x-ray absorption fine structures (EXAFS) measurements were performed at the 1W1B beamline at Beijing Synchroton Radiation Facility (BSRF). The dc magnetization measurements were carried out on a superconducting quantum interference device (SQUID, Quantum Design MPMS XL-7). Magnetization temperature (MT) curves were measured in field-cooled (FC) and zero-field-cooled (ZFC) modes under various magnetic field in the temperature ranges of 5 ~ 400 K. The specific heat against temperature was obtained by using the heat capacity unit on

Physical Property Measurement System (PPMS). Magnetostriction as a function of applied field and temperature were obtained using the strain gauge method on PPMS.

**3 Result and discussions**

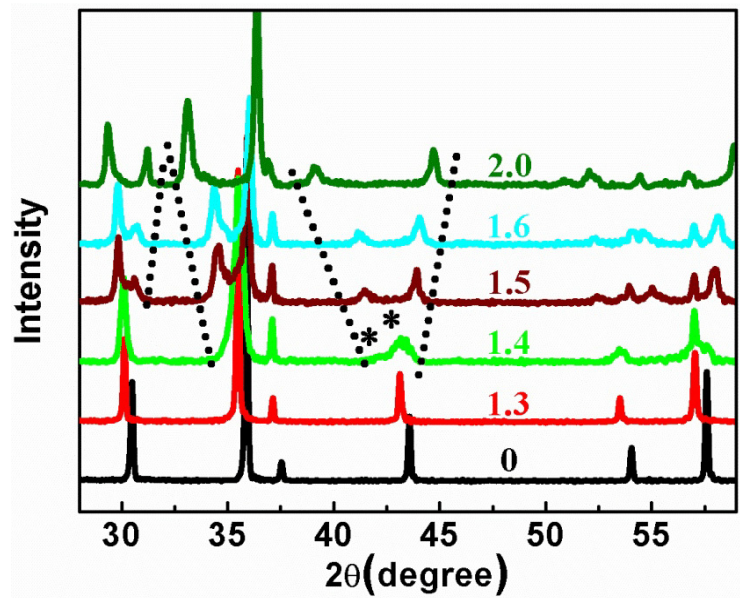

Figure 1. The XRD patterns of the samples with x=0 ~ 2.0, the dash lines are guides for the eye of the movements of the new peaks generated during the structure transition while the star marks present the split of the peak.

The XRD patterns of the series are shown in figure 1. There is no extra peak of second phase within the resolution of XRD method, suggesting a pure quality of the samples studied in this work. For the samples with x < 1.4, it can be well indexed in a cubic spinel structure with space group of *Fd-3m*, as that of $CoCr_2O_4$ (x = 0). While x = 1.4, the peaks become diffusion and some of them start to split, as the one at 43° with star marks. This is caused by the distortion of the crystal lattice induced by doping. When x=1.5, there generates a bunch of new peaks which move along different directions as the increase of dopants, indicated by the dot lines. It is thus reasonable to conclude that there is a structural transition around this composition. This new structure can be defined as tetragonal phase in a space group of *I41/amd*.[23] The diffusion of the peaks for the patterns of high-level-doping samples indicates the poor crystallinity and small grain size of these materials, which is due to the decrease of sintering temperature.

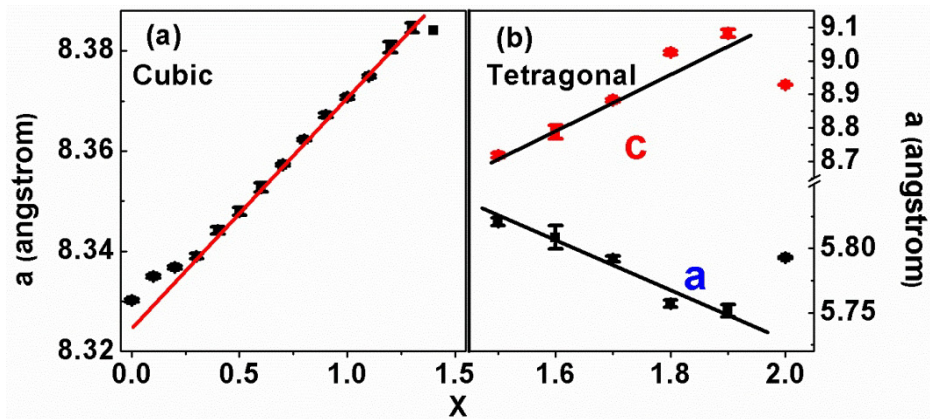

Figure 2. (a)The composition dependence of lattice parameters for the samples with x=0~1.4 which are indexed in a cubic symmetry; (b) shows that of samples with x=1.5~2.0 which are indexed using the tetragonal symmetry.

The x dependence of lattice parameter for the samples with cubic structure is shown in figure 2(a). These parameters were obtained by refining the patterns to the reliability factor $R$p in the range of 6.5 ~ 11.3%. Parameters of the samples with x < 1.4 linearly increase with the increase of x, except for those of x<0.3 which deviate from the fitting line. The overall increase is due to the larger ionic radii of $Mn^{3+}$ in octahedron (0.785 Å) than that of $Cr^{3+}$ (0.755 Å). Meanwhile, the different increment, when x < 0.3 or > 0.3, implies a different site-preference for the doped Mn ions. Since $Co^{2+}$ has the radii of 0.885 Å in octahedral site and 0.75 Å in tetrahedral site, if the doped $Mn^{3+}$ ions firstly replace the $Co^{2+}$ ions on tetrahedral site and make the latter one to take the octahedral sites, the total increment of the lattice parameter will be larger than that of the direct substitution of $Cr^{3+}$ by $Mn^{3+}$. Therefore, one could conclude that the doped Mn ions firstly occupy the Co ions on A-sites ($Co_A$) when x < 0.3, and then mainly take the $Cr_B$-sites while x > 0.3. This interesting change of site-preference during the doping process has also been found in Fe doped $CoCr_2O_4$ system.[22] The lattice parameters of the samples with x > 1.4 are ploted in figure 2(b). One can see that, as the increase of dopants, the lattice is linearly shrinked in the *ab* plane and expanded along the *c* direction. However, the volumes of the samples are still expanded with increased dopants.

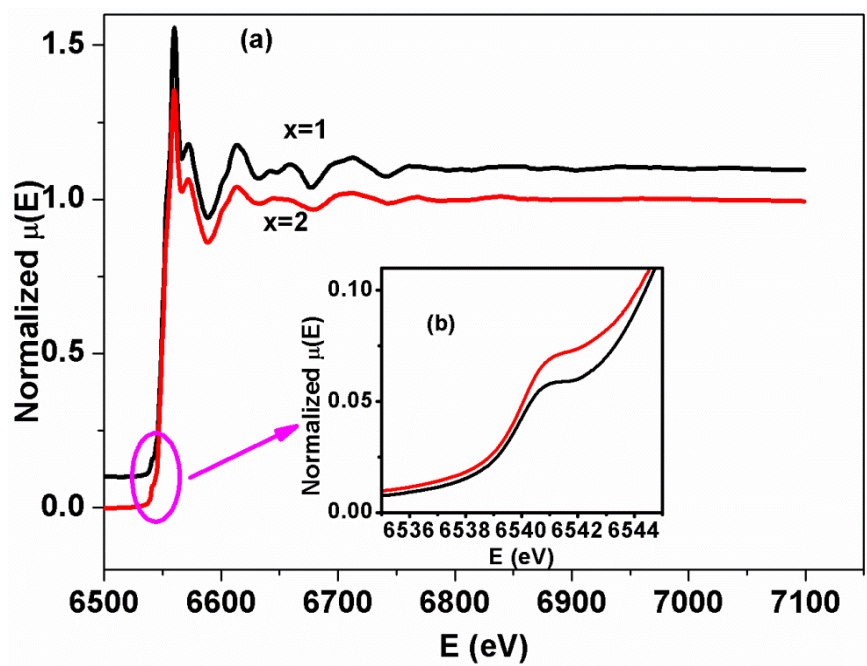

Figure 3. (a)The EXAFS patterns of the Mn K-edge of the samples with x=1.0 and 2.0; (b) the near edge structure of the spectrums.

To further support the above conclusion about site-preference in this system, EXAFS experiment has been carried out in this work to identify the distribution of the doped Mn ions. As shown in figure 3(a), the spectrum of the sample with x = 1.0 is identical with that of x = 2 ($CoMn_2O_4$), indicating a similar ion configuration in them. Since the Mn ions in $CoMn_2O_4$ mainly occupy the B-sites [23,24], it is reasonable to infer that the doped Mn ions in $CoCr_2O_4$ should be mostly distributed on B-sites, as inferred by figure 2. However, figure 3(b) shows that there is a pre-edge peak on both the spectrums. This is due to, according to the previous reports [25,26], Mn ions appeared on the tetrahedral A-sites. Therefore, there should be a small amount, due to the low intensity of the pre-edge peaks, of Mn ions on the A-sites during the doping process. This is identical with the conclusion of the

XRD experiment. Figure 3 also shows that the Mn K-edge threshold energies of the samples are barely moved with the increase of dopants, which means that the valences of the Mn ions in these samples are almost unchanged, about +III as it in $CoMn_2O_4$.[25,26]

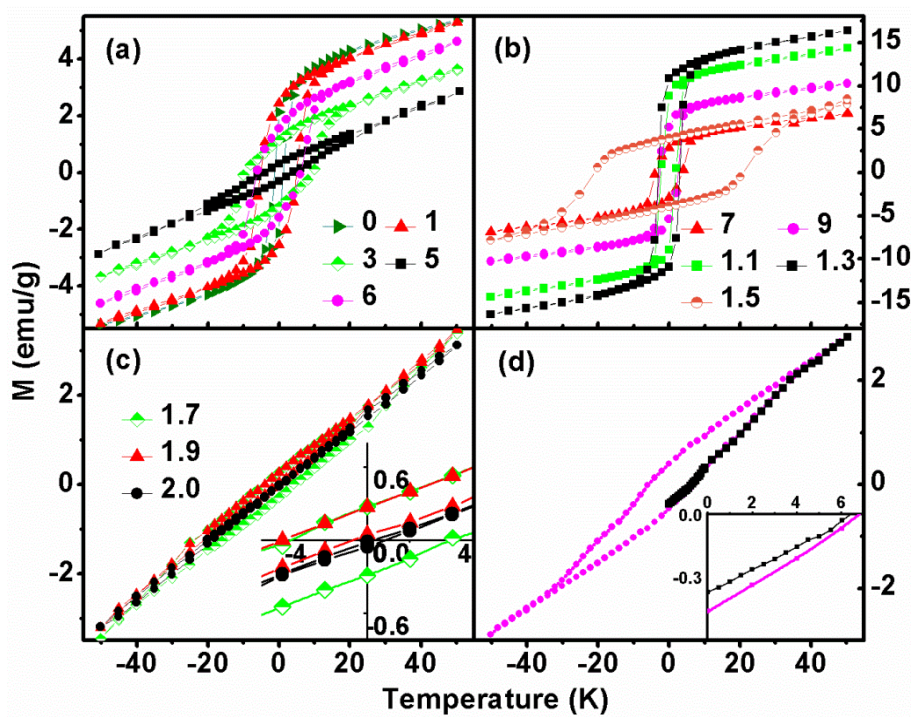

Figure 4. (a), (b) and (c) show the hysteresis loops (MH loop) for the series measured at 5 K. The inset of (c) gives the enlarged view of the MH loops of the samples with x= 1.7, 1.9 and 2.0. (d) shows the MH loop of the sample with x=0.5, measured after field cooling the sample to 5 K under 1 kOe, where the inset shows the fourth quadrant of the loop.

Figure 4(a)-(c) present the MH loops of the series measured at 5 K. One can see that the loops of these samples cannot be fully saturated even at 50 kOe, which is due to the non-collinear magnetic structure in the system. Nevertheless, the high field range of the MH loops becomes more flat with the increase of dopants, as shown in figure 4(b). This indicates the increase of ferromagnetic interactions and thus the change of cone angle of the spiral order in this system. While x > 1.3, the slope of the high field range on MH loops increases again, as shown in figure 4(b), which implies that the distortion of lattice during the structure transition may suppress the ferromagnetic interactions induced by Mn ions and make the system more frustrated. The almost linear MH loops in figure 4(c) shows that the samples with x near 2.0 ($CoMn_2O_4$) have a more frustrated ferrimagnetic structure, which is consist with the previous report.[27] Additionally, an interesting phenomenon has been observed in the MH loop of x=0.5 measured after field cooling the sample to 5 K under 1 kOe, as shown in Figure 3(d). The initial magnetization of the MH loop is negative in the field range of 0~8 kOe. This behavior is attributed to the FC process and will be discussed later.

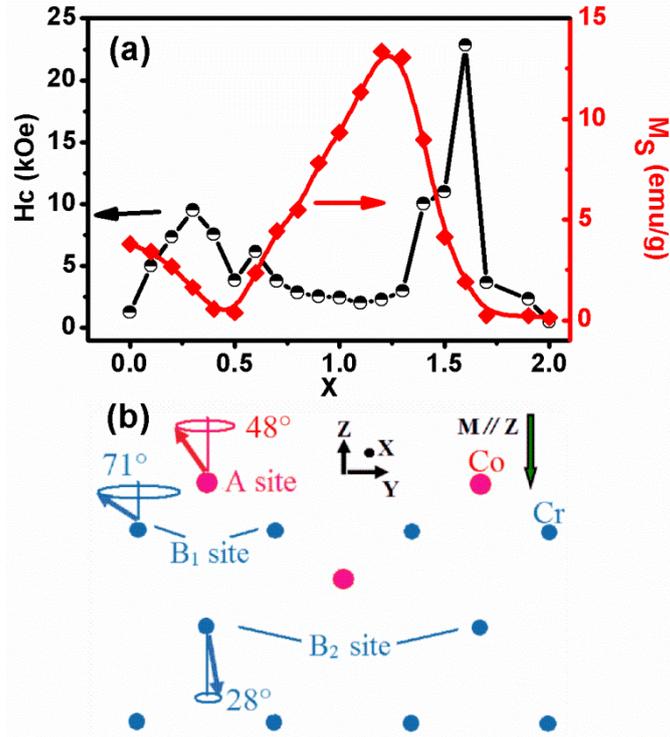

Figure 5. (a) Composition dependences of $M_S$ and $H_C$ measured at 5 K; (b) is the sketch map of the cation distribution and spin configure in $CoCr_2O_4$ viewed from the [100] direction.

Figure 5(a) shows the composition dependence of spontaneous magnetization ($M_S$, defined as the magnetization extrapolated to zero-field from the high field linear region of MH loops) and coercivity ($H_C$, defined as the average value of the negative and positive fields when M = 0 on MH loops) of the series. It can be seen that $M_s$ possesses a minimum close to zero at x=0.5, indicating a composition induced compensation effect, even though the exact compensation point has not been found in this series. When the structure transition starts at around x = 1.4, $M_S$ decreases abruptly to almost zero at x=2.0, which is consist with the result of increased frustration in this system concluded above.

To illustrate this compensation phenomenon, analysis of the magnetic structure in the system is necessary. Figure 5(b) shows the schematic arrangement of the magnetic structure in $CoCr_2O_4$ projected on the YZ plane[12,14]. The moments on A and $B_1/B_2$ sites are conically ordered and their cone axes are antiparallelly aligned to each other. Therefore, the ferrimagnetic components of the conical spins along the cone axes induce a macroscopic moment in the z [001] direction. Considering the spin contributions to magnetic moments of $Mn^{3+}$(4.9 $\mu_B$), $Cr^{3+}$(3.87 $\mu_B$) and $Co^{2+}$(3.87 $\mu_B$) with spin S = 4/2, 3/2 and 3/2, respectively, only if the $B_1$ and A sites are taken by the doped Mn ions, the net magnetic moment of the system will decrease and lead to the observed magnetic compensation. Combine with the conclusion of EXAFS, it is believed that the dominant occupation of Mn ions should be on $B_1$ sites. This suggests that the main contribution to magnetization of the series comes from $B_2$ sites when x < 0.5 and turns to the A+$B_1$ sites while x > 0.5.

Figure 5(a) shows that $H_C$ firstly increases to a value of about 9502 Oe before abruptly decreases to a minimum at x = 0.5. The increment of $H_C$ before x = 0.3 implies that introduction of $Mn^{3+}$ may have enhance the anisotropy in this system, due to the Jahn–Teller effect of $Mn^{3+}$ ions on the tetrahedral A-sites.[27] However, the exchange bias effect (EB), which will be discussed later, suggests the possible Mn-clusters may also have contribution to the increase of $H_C$. The decrease of $H_C$ near x = 0.5 is derived from the abnormal shape of MH loops, as shown in figure 4(a), which has a nearly zero magnetization when H = 0.[22,28,29]. When x > 0.5, where the doped A+$B_1$ sublattices become domination and Mn ions mainly take the $B_1$-sites, $H_C$ decreases to a stable value of about 2000 Oe. This is caused by the increase of ferromagnetic interactions as Mn doping. When the structure transition starts at x = 1.4, $H_C$ begins to increase sharply and reaches a maximum of 22828 Oe at x = 1.6, then it reduces to a relatively low value till x = 2.0. This enormously change of the $H_C$ is clearly due to the variation of anisotropy caused by the lattice distortion during the structure transition.

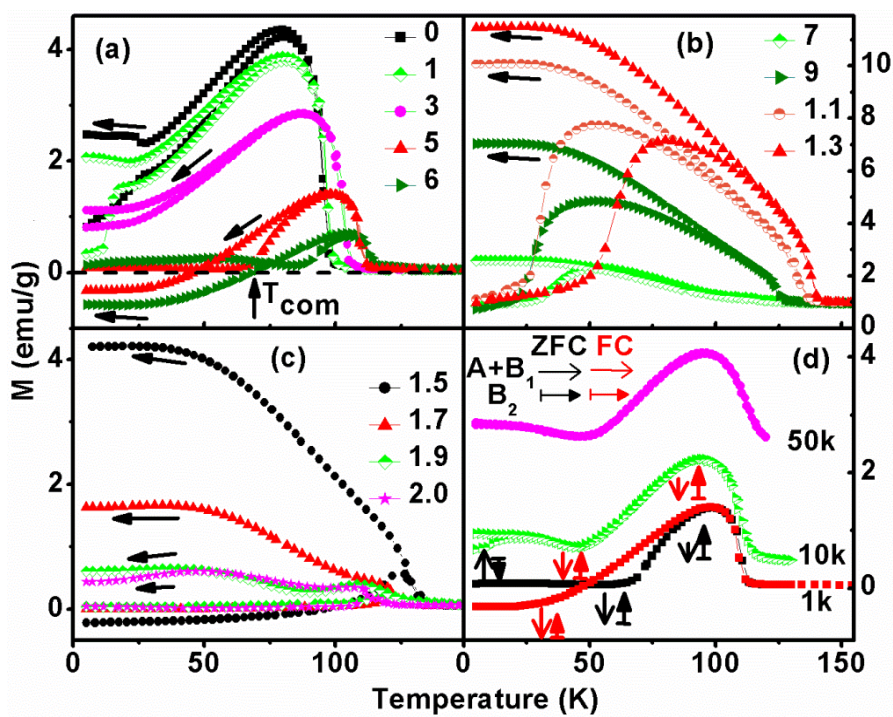

Figure 6. (a), (b) and (c) show the ZFC and FC curves for the series under 1 kOe, while the arrows indicate the FC direction. The ZFC-FC curves of the sample with x=0.5 measured at different field have been shown in (d) where the compensation process has also been presented: the simple arrows indicate the magnetic moments on A and $B_1$ sites for ZFC (black) and FC (red) process, respectively, while the solid arrows present the moments on $B_2$ sites in the same processes.

Figure 6(a)-(c) present the ZFC-FC curves of the samples measured under 1 kOe. One can see that the magnetoelectric transition at 26 K in $CoCr_2O_4$ has been strongly suppressed by Mn doping and becomes unclear in the sample with x = 0.3. This is induced by the enhanced ferromagnetic interactions in the samples. The samples with x = 0.5 and 0.6 show the unusually negative magnetization on the FC curves and a compensation point indicated by the arrow ($T_{com}$). This explains the negative initial magnetization of the MH loop for the sample with x = 0.5, shown in figure 4(d). As the increase of dopants, the bifurcation of ZFC and FC curves becomes more obviously and moves to higher temperature. This is a typical reentrant-spin-glass-like behavior as in $CoCr_2O_4$[14], caused by the possible Mn-clusters in the samples or by the distortion induced by the structure transition.

The ZFC-FC curves of the sample with x = 0.5 measured under different fields are shown in figure 6(d). One can see that the magnetization of FC curve becomes negative below 45K under 1kOe. The external field used during the measurement is lower than $H_C$, as shown in figure 5(a), and thus the energy barrier of anisotropy could not be overcomed by this field. Therefore, there should be no response of the domain walls during the process. Consequently, the negative value of magnetization in this sample implies a role conversion for magnetic contributor at 45K, suggesting a temperature compensation behavior. The mechanism for the compensation effect has been illustrated in figure 6(d), considering the moments on the two magnetic sublattices. In the samples with x = 0.5 and 0.6, the A+B$_1$ sublattice is domination at low temperature while the B$_2$ sublattice becomes domination when the temperature is higher than $T_{com}$. Therefore, the compensation happens at a certain temperature when the moments on the two sublattices become equal to each other. Notice that the ZFC curve of the sample with x = 0.5 shows a flat state through $T_{com}$. This is due to the freeze of domains in a high anisotropy system in which the alignment of the moments by external field becomes very difficult and the role conversion for magnetic contributor is not apparent. During the FC process, the external field cannot parallels the moments on A+B$_1$ sublattice, though it becomes larger than that of B$_2$ sublattice, and thus appears the negative magnetization below $T_{com}$. Additionally, either when x < 0.5 or > 0.6, there is only one dominating sublattice in the whole temperature range from 5 K to 300 K, and thus there is no compensation effect observed.

When the applied field is increased to 10 or 50kOe, which is larger than $H_C$, the MT curve shows a minimum at $T_{com}$ with finite value. This transformation from a compensation point to a non-zero minimum strongly suggests that the magnetic structure has been changed by this high field, most likely the cone angles of the two magnetic sublattices are strongly changed under this field, as the discussion in figure 4. There is a thermal hysteresis between the minimums on ZFC and FC curves measured under low field, which is disappeared at high field. This is due to the pinning of the domain walls, since the domains cannot be switched completely by a field lower than $H_C$.

Table 1. $\Theta_{CW}$, $T_C$ and $f$ resulted from fitting the 1/χ versus T data to the Curie–Weiss description. $\Theta_{CW}$ is taken as the x-intercept of the inverse susceptibility curve, $T_C$ is chosen as the peak point in the first derivative of the susceptibility while $f$ is defined as the absolute ratio of $\Theta_{CW}$ to $T_C$.

| x | $\Theta_{CW}$ (K) | $T_C$ (K) | f |
|---|---|---|---|
| CoCr$_2$O$_4$ | -450 | 94 | 4.7 |
| 0.1 | -416 | 102 | 4.0 |
| 0.3 | -470 | 108 | 4.3 |
| 0.5 | -561 | 115 | 4.8 |
| 0.7 | -548 | 116 | 4.7 |
| 0.9 | -480 | 123 | 3.9 |
| 1.1 | -531 | 132 | 4.0 |
| 1.3 | -436 | 136 | 3.2 |
| 1.5 | -676 | 127 | 5.3 |
| 1.7 | -664 | 126 | 5.2 |
| 1.9 | -652 | 120 | 5.4 |
| CoMn$_2$O$_4$ | -597 | 114 | 5.2 |

Table 1 shows the experimental Curie temperature ($T_C$), paramagnetic Curie-Weiss temperature ($\Theta_{CW}$), and the frustration parameter (f) of the series, which are obtained by fitting the high temperature (300 – 400 K) susceptibility of the samples with the Curie–Weiss equation, χ = C / (χ - $\Theta_{CW}$). The $T_C$ enhances monotonously with the increase of x when x < 1.3. Since the volume of lattice is overall increased with doping, the enhanced $T_C$ thus implies a strengthening of the exchange interactions which could be induced by the increased magnetic moments and the decrease of frustration degree. It suddenly begins to decrease at x = 1.4, which is due to the weak exchange interactions caused by the lattice distortion from cubic to tetrahedral symmetry. $T_C$ of the final compound, CoMn$_2$O$_4$, is 114 K and in accordance with the references.[27,30]

The $f$ factor of CoCr$_2$O$_4$ (4.83) is smaller than the previous report[31], which is due to the different $T_C$ values obtained by different methods. One can see that $\Theta_{CW}$ is overall decreased by the doped Mn ions while x < 1.3, implying the strengthening of ferromagnetic interactions between A-B sites. Consequently, the degree of frustration in the system is also decreased with the increased dopants, as can be seen from the $f$ factors. This is coincident with the discussions above. When the structure transition begins at around x = 1.4, both $\Theta_{CW}$ and $f$ show a sharp increase, which reflects the large frustration in the tetrahedral system. It is thus reasonable to conclude that Mn ions on B-sites could induce a less frustrated magnetic structure in CoCr$_2$O$_4$ until that the lattice distortion counteracts this effect and restore the frustration.

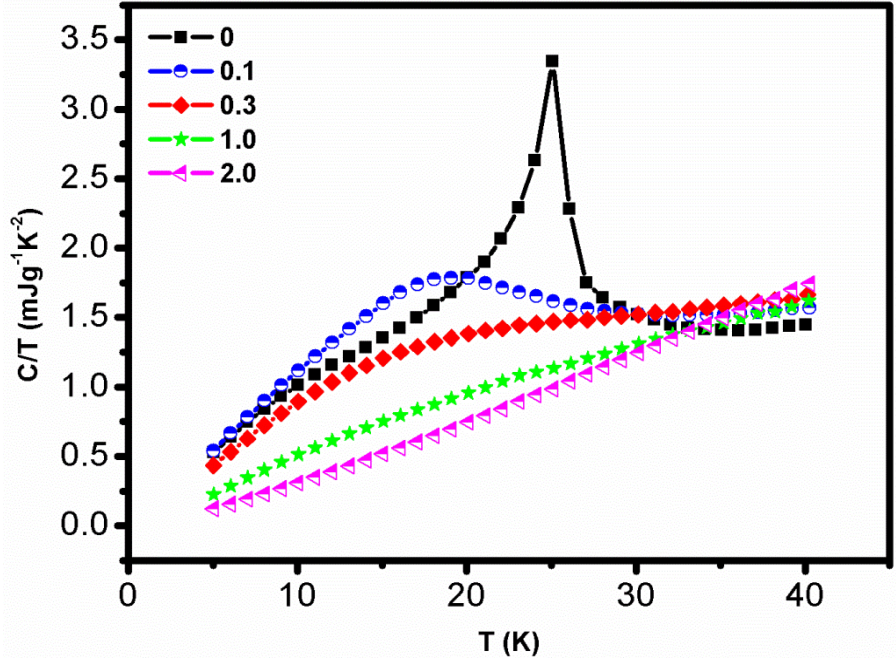

Figure 7. Specific heat over temperature curves of the series from 5 K to 40 K.

Figure 7 gives the specific heat against temperature ($C/T$) for several samples, which could be used as the identification for the magnetoelectric transition in this system and other spinel materials[32]. As the increase of Mn ions, the C/T peak at $T_S$ is dramatically weakened and moves to the lower temperature, indicating that doping on *B*-sites could effectively suppress the conical spin order in the system, as shown in figure 6. This should be due to the Mn$^{3+}$ ions on B-sites which break the long range spiral order and also strengthen the ferromagnetic exchange interactions in the system. The low intensity and wide temperature range of the C/T peaks for the samples with increased dopants imply a diffusion of the magnetic transition and unsaturation of the coherent length of conical orders. At last, the magnetoelectric transition is fully suppressed above x=0.3.

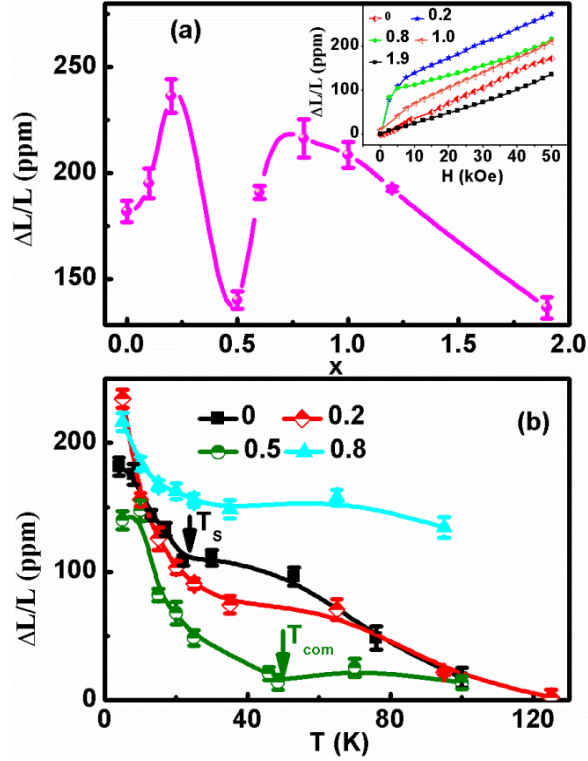

Figure 8. (a) The composition dependence of magnetostriction under 5T at 5K. The inset shows the curves of strain against external field for different samples. (b) The temperature dependences of magnetostriction under 5T of the series.

Figure 8(a) gives the composition dependence of magnetostriction for the series under 5 K, which had not been investigated in this system so far, to the best of our knowledge. It has been found that the $CoCr_{2-x}Mn_xO_4$ system exhibits a quite large magnetostrictive, value from 139 to 240 ppm. As shown in figure 8(a), the magnetostriction shows a maximum at x = 0.20 and a minimum at about 0.50. The increase of strain before x=0.2 is related to the occupation of $Mn^{3+}$ ions on A sites, due to their Jahn–Teller effect on the tetrahedral sites mentioned above[27]. The Co ions which are forced to take the octahedral B-sites may also lead to the increase of magnetostriction as the situation in $CoFe_2O_4$[33]. The decrease of strain approaching x = 0.5 should be caused by the reduction of coupling between magnetization and crystal lattice near this compensation point. The inset of figure 8(a) gives the field dependence of the strain (ΔL/L) for several samples. It clearly indicates that there is a magnetostriction existed natively in $CoCr_2O_4$, which should derives from the conical spin structure in the system. However, the H dependence of the strain does not have the same style with the MH behavior, which needs further investigations. The doped $Mn^{3+}$ ions sharply enhance the low-field slope and the magnitude of strain curves, which could be attributed to the introduction of new source of magnetostriction, as mentioned above. While x ≥ 1.0, the strain turns to linear shape again and its value starts to decrease, implying the weak coupling between magnetization and lattice due to the suppressed conical order in this system.

The temperature dependences of strain under 5 T of the series are shown in figure 8(b). The curves with x=0 and 0.5 show the similar characteristic as the MT curves at the magnetoferroic transition and compensation temperatures. At temperature below $T_S$, the large magnetostriction implies a strong coupling between the conical order and the crystal lattice, which gradually decreases as the increase of temperature. The interesting relationship between this magnetostriction effect and the magnetoelectric property in this temperature range suggests a similar origin of these two characteristics. Since the long-range spin order collapses above $T_S$, the strain continuously reduces with the increasing of temperature. Interestingly, the magnetostriction also shows a minimum at $T_{com}$, implying that the stain below and above this temperature is derived from two different magnetic sublattices as discussed above. Though there is no magnetoelectric transition has been observed in the other samples, the slope of their strain curves still changes with temperature, which reveals a developing of local conical spin orders in these systems. And thus there may generate a glass state in this system, as mentioned in figure 6. It is noteworthy that the sample with x = 0.8 maintains a high value of strain in a large temperature range, indicating a strong and robust coupling between the magnetism and the crystal lattice in this sample.

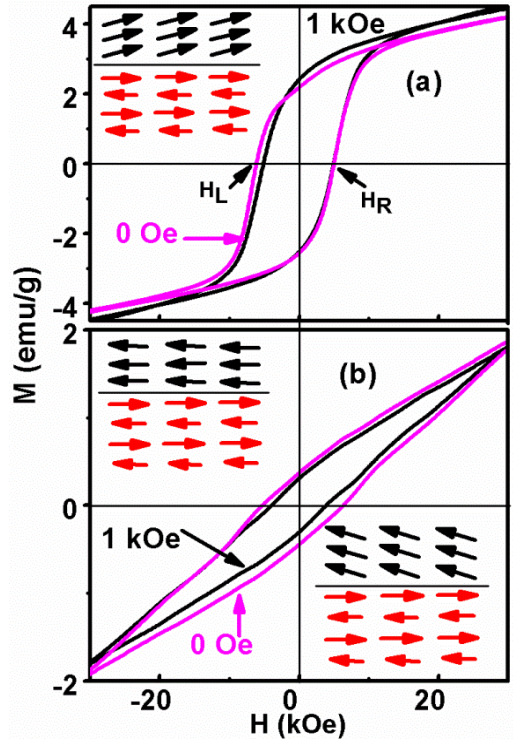

Figure 9. The MH loops measured under different FC processes (0 Oe and 1 kOe) of the samples with (a) x=0.1 and (b) 0.5, respectively. The insets show the schematic diagram of the spin configurations at the interface of the Mn clusters and the matrix.

The series studied in this work have also possessed an EB effect in the composition range of x=0.1~0.5. It is well known that materials with coupling between ferromagnetic and antiferromagnetic phases can exhibit an exchange bias after undergoing a FC process[34], which is attributed to the unidirectional anisotropy due to the exchange coupling between them.[35,36] However, there is no obvious second phase can be detected in this series. Therefore, based on the discussions about the distribution of Mn ions above, we've concluded that there should be some ferromagnetic (FM) Mn clusters in the A+$B_1$ sublattice, generated by the unavoidable inhomogeneity of the samples or the local short-range spiral order mentioned above. And thus we present the following interpretation for the EB effect in this work. As can be seen from figure 9(a), the exchange bias field is defined as $H_{EB}=(H_R-H_L)/2$, where $H_L$ and $H_R$ are the negative and positive coercive fields when magnetization equals zero, respectively. Since the matrix is ferromagnetic (FI) and the Mn-clusters are FM, the pinning phase should be the matrix which will be ordered firstly by the external field during the FC process. For the sample with x=0.1, the moments of Mn clusters could be easily parallelled by the external field during the FC process and thus leads to a normal EB effect, as shown in the inset of figure 9(a). However, for the sample with x=0.5, the antiferromagnetic interactions between the Mn clusters and the matrix are too strong to be overcome by the external field during the FC process. Thus, the moments on the interface between them will be still antiferromagneticly aligned, as shown in the inset of figure 9(b). Consequently, the pinning effect will happen in the fourth quadrant of the MH loop, and induces an unusual negative $H_{EB}$ in the system.

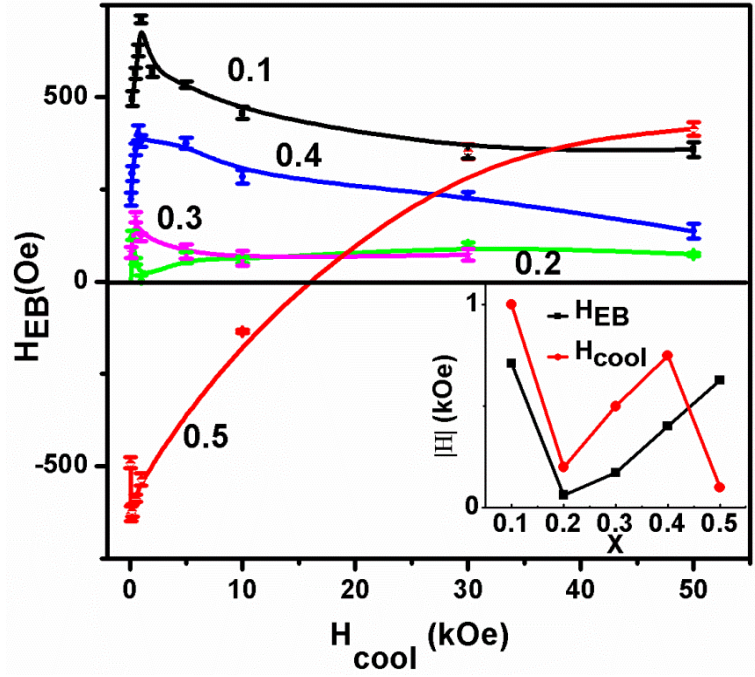

Figure 10. $H_{cool}$ dependence of $H_{EB}$ for different samples, the inset shows the composition dependences of $|H_{EB}|$ and $H_{cool}$.

The cooling field ($H_{cool}$) dependence of $H_{EB}$ for the samples with x=0.1 ~ 0.5 is shown in figure 10. It is similar with a typical EB system, showing a maximum value at a certain $H_{cool}$. Especially, $H_{EB}$ of the sample with x = 0.5 becomes positive ($H_R < H_L$) when $H_{cool}$ becomes larger than 10 kOe, which means that this external field can overcome the anisotropy at the interface and make the two phases parallel aligned during the FC process. It is noteworthy that this field is also larger than the field under which the FC curve of x = 0.5 becomes positive, as shown in figure 6, and thus suggests that the temperature compensation and EB effect in the system have different origins.

The inset of figure 10 shows the composition dependences of the absolute value of $H_{EB}$ ($|H_{EB}|$) and $H_{cool}$. $|H_{EB}|$ shows a minimum at x = 0.2, which implies the different origins of EB on the two sides of this component. As the increase of dopants, the proportion of FM clusters grows continuously which will certainly reduce the $H_{EB}$[37]. While x > 0.2, the Mn ions start to occupy B1-sites and create new form of FM clusters in the ferrimagnetic matrix, which causes the revival of EB until x = 0.4 whereafter the $H_{EB}$ begins to decrease again.

**4 Conclusion**

In conclusion, the magnetic compensation, magnetostriction and exchange bias of $CoCr_{2-x}Mn_xO_4$ (x = 0 ~ 2) series have been studied. We find out that the doped $Mn^{3+}$ ions start to occupy A-sites initially when x < 0.2, and then prefer to take B1-sites. This results in a role conversion of magnetic contributors and a composition compensation between two competitively magnetic sublattices near x = 0.5. In the samples with x = 0.5 and 0.6, a temperature dependence compensation can be found in the temperature range of 45 ~ 75 K. The long range conical spin order is suppressed by doping, with $T_S$ shifts to low temperature and the C/T peak becomes weak. For the first time, the magnetostriction effect with a strain of 240 ppm has been observed in the sample with x = 0.2 at 5 K. It has also been found that the magnetostrictive property is identical with the composition and temperature compensation behavior. Furthermore, the doped Mn ions form FM clusters on both A and B1 sites, which leads to the occurrence of exchange bias in this system and a negative $H_{EB}$ for x = 0.5.

**Acknowledgement**

This work was supported by the National Basic Research Program (2009CB929501), specific funding of Discipline & Graduate Education Project of Beijing Municipal Commission of Education, and the National Natural Science Foundation of China in Grant No. 11174352.

Physical Review B 77 (2008) 094412.